\documentclass[12pt,draftcls,onecolumn]{IEEEtran}

\usepackage[dvips]{color}
\usepackage{graphicx}
\usepackage{amsmath}

\begin{document}

\title{Relay Selection for Wireless Communications Against Eavesdropping: A Security-Reliability Tradeoff Perspective}

\markboth{IEEE Network (accepted to appear)}%
{Yulong Zou \MakeLowercase{\textit{et al.}}: Relay Selection for Wireless Communications Against Eavesdropping: A Security-Reliability Tradeoff Perspective}

\author{Yulong~Zou,~\IEEEmembership{Senior Member,~IEEE,}
        Jia~Zhu, Xuelong~Li,~\IEEEmembership{Fellow,~IEEE,}
        and Lajos~Hanzo,~\IEEEmembership{Fellow,~IEEE}

\thanks{Y. Zou and J. Zhu are with the School of Telecommunications and Information Engineering, Nanjing University of Posts and Telecomm., P. R. China. (Email: \{yulong.zou, jiazhu\}@njupt.edu.cn)}
\thanks{X. Li is with the Center for OPTical IMagery Analysis and Learning (OPTIMAL), Xi'an Institute of Optics and Precision Mechanics, Chinese Academy of Sciences, Xi'an 710119, Shaanxi, P. R. China. (Email: xuelong\_li@opt.ac.cn)}
\thanks{L. Hanzo is with the University of Southampton, Southampton, United Kingdom. (Email: lh@ecs.soton.ac.uk)}

}

\maketitle

\vspace{-0.55 in}

\begin{abstract}
This article examines {{the secrecy coding aided}} wireless communications from a source to a destination in the presence of an eavesdropper from a security-reliability tradeoff (SRT) perspective. Explicitly, the security is quantified in terms of the intercept probability experienced at the eavesdropper, while the outage probability encountered at the destination is used to measure the transmission reliability. We characterize the SRT of conventional direct transmission from the source to the destination and show that if the outage probability is increased, the intercept probability decreases, and vice versa. We first demonstrate that the employment of relay nodes for assisting the source-destination transmissions is capable of defending against eavesdropping, followed by quantifying the benefits of single-relay selection (SRS) as well as of multi-relay selection (MRS) schemes. More specifically, in the SRS scheme, only the single ``best" relay is selected for forwarding the source signal to the destination, whereas the MRS scheme allows multiple relays to participate in this process. It is illustrated that both the SRS and MRS schemes achieve a better SRT than the conventional direct transmission, especially upon increasing the number of relays. Numerical results also show that as expected, the MRS outperforms the SRS in terms of its SRT. Additionally, we present some open challenges and future directions for the wireless relay aided physical-layer security.

\end{abstract}

\vspace{-0.15 in}


\IEEEpeerreviewmaketitle

\section{Introduction}

Recently, physical-layer security (PLS) has attracted an increasing research attention [1]-[3], since it was shown to achieve perfect resilience against eavesdropping attacks. The PLS work was established by Wyner [4] for a discrete memoryless wiretap channel comprised of a source and a destination in the presence of an eavesdropper. It was shown in [4] that simultaneous secure and reliable transmissions can be achieved without using secret keys. In [5], Leung-Yan-Cheong and Hellman examined the Gaussian wiretap channel and introduced the notion of \emph{secrecy capacity}, which is derived as the difference between the capacity of the main link spanning from the source (S) to the destination (D) and that of the wiretap link spanning from S to the eavesdropper (E). However, the secrecy capacity of wireless communications is severely affected by the time-varying multipath fading [6]. For example, if the wiretap link has a relatively good condition while the main link experiences shadow fading, the wireless secrecy capacity would drop dramatically. More explicitly, assuming that the S-E link has a certain channel capacity, but the S-D link is severely faded and hence has a low capacity, would increase the probability of E intercepting the legitimate transmissions.

To this end, extensive research efforts have been devoted to enhancing the wireless secrecy capacity in the face of multipath fading for example by invoking diverse multiple-input multiple-output (MIMO) schemes [7], beamforming [8], [9] and cooperative relaying [10], [11]. In [7], Goel and Negi considered the employment of MIMOs for generating a specifically-designed artificial noise to confuse the eavesdropper. It was shown in [7] that the number of antennas of the legitimate transmitter should be higher than that of the eavesdropper for the sake of ensuring that the artificial noise only impacts the eavesdropper adversely without affecting the legitimate receiver. Further beamforming techniques were studied in [8] and [9], which enable the source to transmit its signal in a particular direction to the legitimate receiver, so that the signal arriving at the eavesdropper encounters destructive interference and becomes much weaker than that received at the legitimate receiver experiencing constructive interference, hence leading to a significant secrecy capacity improvement. Additionally, in [10] and [11], we studied the use of relays for guarding against eavesdropping and proposed the single ``best" relay selection technique for enhancing the wireless secrecy capacity.

The aforementioned contributions are mainly focused on improving the wireless security without paying much attention to the communication reliability. To this end, in [12], we investigated the security-reliability tradeoff (SRT) encountered in wireless communications without using any secrecy coding, where the security is quantified in terms of the probability that E succeeds in intercepting the source signal, while the reliability represents the probability that an outage event is encountered at the legitimate destination. These probabilities are termed as the intercept probability (IP) and outage probability (OP), respectively, where the OP can be reduced upon increasing the transmit power of S, but at the same time this also enhances the S-E channel capacity and increases the IP. It was shown mathematically in [12] that upon increasing the IP, the OP is reduced and vice versa, which indicates a tradeoff between the security and reliability. Furthermore, we proposed the single best-relay selection scheme in [12] for achieving a SRT enhancement and showed that as the number of relays increases, the wireless SRT significantly improves. {{It has to be pointed out that the SRT studied in [12] is based on the assumption that no secrecy coding is used.}} However, in the recent literature on PLS [13], more and more secrecy coding techniques have been devised.

As a consequence, this article investigates the SRT benefits of secrecy coding aided wireless communications against eavesdropping attacks, {{differing from [12], where no secrecy coding is considered for formulating and evaluating the wireless SRT}}. The main contributions of this article are summarized as follows. First, we introduce a general channel model of secrecy coding based wireless communications and characterize both the wireless security as well as reliability in terms of the IP and OP, respectively. We will show that as the IP is increased i.e. the security degrades, the OP (reliability) improves and vice versa. Secondly, we characterize the benefits of both single-relay selection (SRS) and multi-relay selection (MRS) schemes in terms of their ability to improve the wireless SRT. Specifically, the SRS scheme chooses the single ``best" relay for assisting the transmissions from S to D, whereas in the MRS approach, multiple relays are selected to forward the source transmissions. Additionally, numerical results will be provided for quantifying the advantage of the SRS and MRS over conventional direct transmission in terms of their SRTs.

The remainder of this article is organized as follows. In Section II, we present the system model of a secrecy coding based wireless communication network consisting of a source and a destination in the presence of an eavesdropper. The IP and OP are invoked for characterizing the wireless security and reliability, respectively. Next, in Section III, we show the benefits of relay nodes in terms of assisting the S-D transmissions and introduce both the SRS and MRS schemes for the sake of guarding against eavesdropping, where numerical SRT results are also provided. Section IV presents some challenging issues, which remain open at the time of writing. Finally, we provide some concluding remarks in Section V.

\section{Security-Reliability Tradeoff for Wireless Communications}
\begin{figure*}
  \centering
  {\includegraphics[scale=0.7]{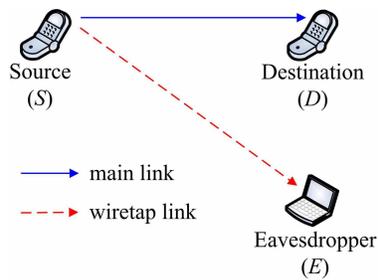}\\
  \caption{A wireless system consisting of a source and a destination in the presence of an eavesdropper.}\label{Fig1}}
\end{figure*}

In wireless networks, the broadcast nature of radio propagation makes the wireless transmission extremely vulnerable to eavesdropping attacks, since it can be readily tapped by an eavesdropper. As shown in Fig. 1, S transmits its signal to D, while E may overhear the legitimate transmission, as long as it lies in the transmit coverage of S. {{Notice that only the single eavesdropper is considered throughout this article and similar results can be obtained for a multi-eavesdropper scenario. It is expected that increasing the number of eavesdroppers would enhance the probability of successfully intercepting the source-destination transmission, resulting in a performance degradation of the wireless security.}} In existing wireless systems, cryptographic techniques are used for preventing E from intercepting the legitimate transmission between S and D. Although the cryptographic methods do indeed improve the transmission security, this comes at the expense of an increased computational complexity and latency. To be specific, a cryptographic algorithm enhances the communication security, but unfortunately requires more computational resources both for encryption as well as for decryption and increases the latency {{[14]}}. Additionally, the encrypted information may still be decrypted by an eavesdropper, for example by using an exhaustive key search known as `brute-force' attack.

To this end, PLS emerges as a promising means of achieving information-theoretic security for confidential communications in the face of eavesdropping. It has been proven in [2] that perfect secrecy becomes possible, when the capacity of the main link spanning from S to D is higher than that of the wiretap link spanning from S to E. Moreover, the capacity difference between the main link and wiretap link was termed as the so-called \emph{secrecy capacity} [5]. To be specific, the secrecy capacity is the theoretic maximum rate at which S can transmit to D in a near-error-free manner and at the same time, without leaking any confidential information to E (i.e. achieving the zero mutual information between S and E). The goal of a secrecy coding algorithm is to make it possible for S and D communicating both reliably and securely.

\begin{figure*}
  \centering
  {\includegraphics[scale=0.8]{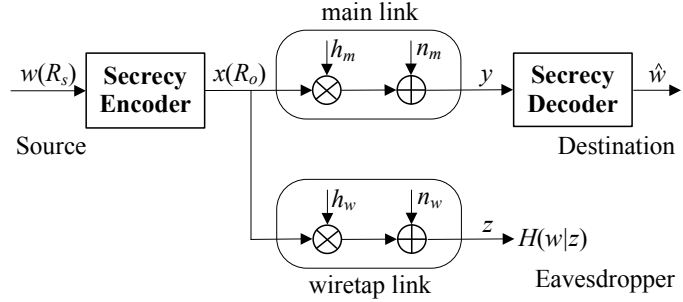}\\
  \caption{A general channel model for secrecy coding based wireless communications.}\label{Fig2}}
\end{figure*}
Recently, an increasing attention has been devoted to the design of practical secrecy coding algorithms (e.g., polar coding [13]) for achieving the information-theoretic secrecy capacity. Fig. 2 depicts a general channel model of secrecy coding based wireless communications, where S intends to transmit its message denoted by $w$ at a secrecy rate of $R_s$. As shown in Fig. 2, the secret message $w$ is first encoded by a secrecy encoder (e.g. polar coding), generating the codeword $x$ at an increased overall rate $R_o$. The rate increment $R_e = R_o - R_s$ represents extra redundancy, which is the cost of providing protection against eavesdropping. Next, the codeword $x$ is transmitted at a power $P$ to D over the main link, which encounters a fading coefficient $h_m$ and an additive white Gaussian noise (AWGN) sample $n_m$. Meanwhile, E also overhears the transmission of S through the wiretap link, where a fading coefficient $h_w$ and an AWGN sample $n_w$ are experienced. Throughout this article, the Rayleigh fading model is considered for both the main and the wiretap links, thus $|h_m|^2$ and $|h_w|^2$ are exponentially distributed with respective means of $\sigma^2_m$ and $\sigma^2_w$. It is also assumed that the AWGN at both S and E has a zero mean and a variance of $N_0$.

\begin{figure*}
  \centering
  {\includegraphics[scale=0.8]{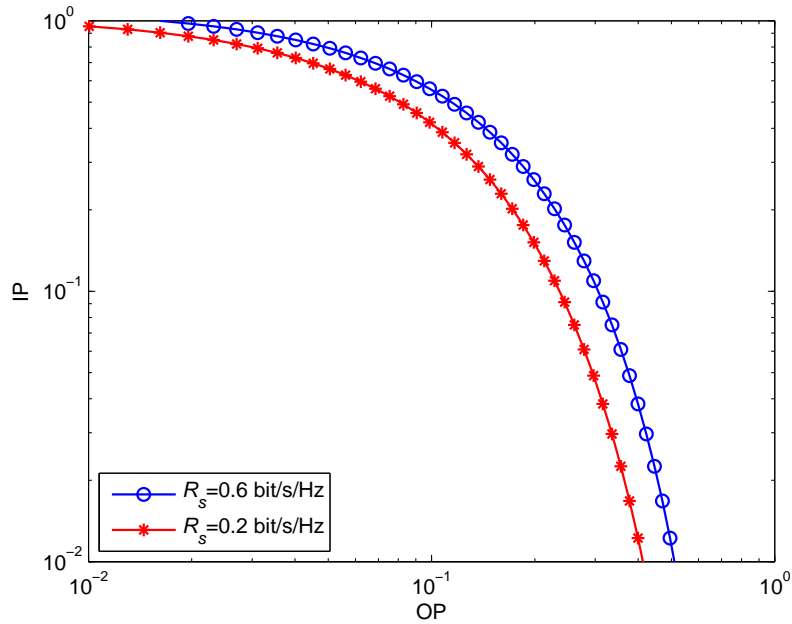}\\
  \caption{IP versus OP of the classic direct transmission scheme for different secrecy rates $R_s$ with {{an SNR of $15{\textrm{ dB}}$}}, $\sigma^2_m=1$, and $\sigma^2_w=0.1$.}\label{Fig3}}
\end{figure*}

According to the Shannon's coding theorem, if the S-D channel capacity drops below the transmission rate {{$R_o$}} (owing to the wireless fading), D fails to recover the source message $x$, hence an outage event occurs. Thus, the OP experienced at D relying on the direct transmission $P_{{\textrm{out}}}^{{\textrm{direct}}}$ can be obtained by calculating the probability of occurrence of an outage event. Additionally, as shown in [15], if the capacity of the wiretap channel becomes higher than the rate increment of {{$R_o-R_s$}}, perfect secrecy is no longer achievable, which is regarded as an event of successfully intercepting the source message, called intercept event. Accordingly, the IP experienced at E with the aid of direct transmission $P_{{\textrm{int}}}^{{\textrm{direct}}}$ is determined by computing the probability of occurrence of an intercept event. Combining the OP and IP expressions, we arrive at ({{see [12] for more information}})
\begin{equation}\label{equa1}
P_{{\textrm{int}}}^{{\textrm{direct}}}  = \exp \left( { - \frac{{2^{ - R_s }N_0  - 2^{ - R_s }\sigma _m^2 P \ln (1 - P_{{\textrm{out}}}^{{\textrm{direct}}} )  - N_0}}{{\sigma _w^2 P }}} \right),
\end{equation}
which characterizes the relationship between the IP and OP for the direct transmission scheme. Fig. 3 shows the IP versus OP by plotting (1) for different secrecy rates $R_s$ with{{ the signal-to-noise ratio (SNR) $P/N_0$ of $15{\textrm{ dB}}$}}, $\sigma^2_m=1$, and $\sigma^2_w=0.1$. It is seen from Fig. 3 that for the secrecy rates of $R_s = 0.2{\textrm{ bit/s/Hz}}$ and $ 0.6{\textrm{ bit/s/Hz}}$, the IP decreases, upon increasing the OP. Again, this implies that the wireless security can be improved at the cost of a reliability degradation and vice versa, explicitly demonstrating the SRT of wireless communications in presence of eavesdropping attacks. One can also observe from Fig. 3 that given the maximum tolerable OP, the IP increases, as the secrecy rate increases from $R_s = 0.2{\textrm{ bit/s/Hz}}$ and $0.6{\textrm{ bit/s/Hz}}$. Conversely, given a target IP requirement, upon increasing the secrecy rate, the OP increases accordingly, demonstrating the SRT degradation imposed when a higher secrecy rate is used in wireless systems.

\section{Relay Selection for Wireless SRT Improvement}

\begin{figure}
  \centering
  {\includegraphics[scale=0.7]{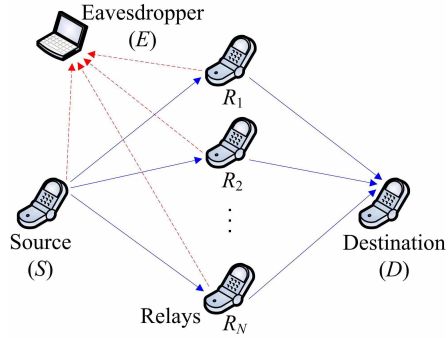}\\
  \caption{A wireless network consisting of multiple relays assisting the S-D transmission in the presence of an eavesdropper.}\label{Fig4}}
\end{figure}
In this section, we consider a wireless relay network, where multiple relay nodes (RNs) are available for assisting the legitimate S-D transmission, as shown in Fig. 4. More specifically, D is assumed to be beyond the coverage area of S, hence $N$ RNs are used for assisting the S-D transmission. Meanwhile, E is assumed to be capable of overhearing the transmissions of both S and RNs, which is the worst-case scenario. For notational convenience, the set of $N$ RNs is denoted by ${\cal{R}} = \{ {R_i}|i = 1,2, \cdots ,N\} $, where the decode-and-forward (DF) protocol is employed by the RNs. Naturally, similar results can also be obtained for the amplify-and-forward (AF) relaying protocol.

Similarly to the channel model of Fig. 2, S first generates its signal $x$ at a secrecy rate $R_s$, which is then transmitted after secrecy coding at an overall rate of $R_o$ to the $N$ relays. Next, the RNs attempt to decode $x$ based on the DF protocol. The specific set of RNs capable of successfully decoding the source signal $x$ is denoted by ${\cal {D}}$, which is termed as the decoding set. Given the $N$ relays, there are $2^N$ possible combinations for the decoding set ${\cal {D}}$, hence the sample space of ${\cal {D}}$ is expressed as $\Omega  = \left\{ {\emptyset ,{\cal {D}}_1 ,{\cal {D}}_2 , \cdots ,{\cal {D}}_n , \cdots ,{\cal {D}}_{2^N  - 1} } \right\}$, where $ \emptyset$ represents the empty set and $ {\cal {D}}_n$ represents the $n{\textrm{-th}}$ non-empty subset of the $N$ relays. If the decoding set is empty (i.e. no RN succeeds in decoding $x$), then all the relays remain silent and D becomes unable to decode the source signal in this case. If the decoding set ${\cal {D}}$ is not empty, we may activate some relays within ${\cal {D}}$ for forwarding the source signal $x$ to D. In what follows, we consider a pair of relay selection approaches, namely the SRS and MRS techniques.

\subsection{SRS Scheme}
In the SRS scheme, only a single RN is selected from the decoding set $\cal{D}$ to assist the S-D transmission. Since E is typically passive and remains silent in wireless networks, in practice it is challenging to estimate the eavesdropper's channel state information (CSI). Motivated by this observation, we assume that only the CSI of the S-D link is used for performing the relay selection, whilst the eavesdropper's CSI knowledge is unavailable. Hence, the specific RN that maximizes the legitimate transmission capacity is typically considered as the ``best" relay for forwarding the source message. Accordingly, the best SRS criterion is formulated as
\begin{equation}\label{equa2}
{\textrm{Best Relay}} = \arg \mathop {\max }\limits_{i \in {\cal {D}}_n } |h_{id} |^2,
\end{equation}
where $h_{id}$ represents the fading coefficient of the channel spanning from the RN $R_i$ to D. It can be observed from the SRS criterion of (2) that only the main channel's CSI is required without the wiretap channel's CSI knowledge. Using the above relay selection, the capacity of the main channel spanning from the ``best" relay to D denoted by $C^{\textrm{single}}_{bd}$ can be easily determined, where the subscript `$b$' stands for the ``best" relay. As discussed above, when D fails to decode the source signal, an outage event is encountered at D. Therefore, using the law of total probability [9], we can obtain the OP of the SRS scheme by calculating the probability that $C^{\textrm{single}}_{bd}$ is less than $R_o$.

When the ``best" relay forwards the source signal to D, it can be overheard by E, as shown in Fig. 4. Meanwhile, E is assumed to be within the source node's transmit coverage, thus it can also overhear the direct transmission from S. Hence, E can combine its signals received from both S and the ``best" relay to obtain an enhanced signal version using selective diversity combining (SDC), equal gain combining (EGC), or maximum ratio combining (MRC). Typically, MRC is capable of achieving a better combining performance than the SDC and EGC. We thus consider the employment of MRC at E for the sake of maximizing its capability of interpreting the source message. Additionally, given the case of ${\cal {D}}=\emptyset$ (i.e. no RN succeeds in decoding the source message), all relays remain silent and thus E can only overhear the direct transmission from S. As mentioned, when the eavesdropper's channel capacity becomes higher than the rate increment of $R_e $, perfect secrecy can no longer be achieved and a so-called intercept event occurs. Hence, the IP of the SRS scheme can be readily determined by comparing the eavesdropper's channel capacity and $R_e$ [15].

\subsection{MRS Scheme}
In contrast to the SRS, the MRS scheme allows multiple relays to simultaneously forward the source signal to D. To be specific, if the decoding set $\cal {D}$ is non-empty (i.e. ${\cal {D}}={\cal {D}}_n$), all the relays in the decoding set of ${\cal {D}}_n$ may be activated for forwarding the source signal to D. This is different from the SRS scheme, where only the single ``best" relay is selected from the decoding set to assist the S-D transmission. In the MRS scheme, a weight vector ${{\textbf{w}}} = [w_1 ,w_2 , \cdots ,w_{|{\cal{D}}_n|} ]^T$ is defined for activating the RNs that succeeded in decoding the source signal, where $(\cdot)^T$ represents the transpose operation and $|{\cal {D}}_n|$ represents the cardinality of the decoding set ${\cal{D}}_n$. Moreover, the total transmit power of all relays should be constrained to unity in order to make a fair comparison in terms of power consumption, hence {{the Euclidean norm of weight vector ${\textbf{w}}$ is constrained to be one}}. In the case of ${\cal {D}}={\cal {D}}_n$, all relays in the decoding set ${\cal {D}}_n$ are activated for simultaneously transmitting the source signal with the aid of the weight vector. We consider that ${{\textbf{w}}}$ is optimized for maximizing the received SNR at D, yielding
\begin{equation}\label{equa3}
\begin{split}
\mathop {\max }\limits_{\textbf{w}} {\textrm{ }} |{\textbf{w}}^T {\textbf{H}}_d |^2 P/N_0 ,
\end{split}
\end{equation}
{{under the condition that the Euclidean norm of weight vector ${\textbf{w}}$ is constrained to be one}}, where ${\textbf{H}}_d  = [h_{1d} ,h_{2d} , \cdots ,h_{|{\cal{D}}_n |d} ]^T$ represents the vector of fading coefficients for the channels spanning from all relays in ${\cal D}_n$ to D. According to the Cauchy-Schwarz inequality, an optimal weight vector ${\textbf{w}}_{{\textrm{opt}}}$ can be readily obtained from (3) as ${\textbf{w}}_{{\textrm{opt}}}  = {{{\textbf{H}}_d^* }}/{{|{\textbf{H}}_d |}}$, which shows that the weight vector optimization only requires the main channel's CSI without the need of the eavesdropper's CSI knowledge. Using the optimal vector and the Shannon capacity formula, we can readily obtain the channel capacity achieved at D, which is then substituted into the outage definition for determining the OP of the MRS scheme.

Due to the broadcast nature of radio propagation, E would overhear the transmissions of all relays in ${\cal D}_n$. Meanwhile, E can also overhear the direct transmission from S, as shown in Fig. 4. Here, the MRC method is considered for E to combine its received signals from both the S and relays. After that, we can obtain an enhanced SNR at E and then determine the channel capacity achieved at E. Finally, the IP experienced at E relying on the MRS scheme can be obtained by calculating the probability that the eavesdropper's channel capacity becomes higher than the rate increment $R_e$.

\subsection{Numerical Comparison}
In this subsection, we present numerical SRT results for the conventional direct transmission as well as the SRS and MRS schemes. In this article, the wireless amplitudes (i.e., $|h_{sd}|$, $|h_{si}|$, $|h_{id}|$, $|h_{se}|$ and $|h_{ie}|$) are modeled by the Rayleigh fading, which in turn, leads to the fact that the squared magnitudes $|h_{sd}|^2$, $|h_{si}|^2$, $|h_{id}|^2$, $|h_{se}|^2$ and $|h_{ie}|^2$ are exponentially distributed random variables with their respective means denoted by $\sigma^2_{sd}$, $\sigma^2_{si}$, $\sigma^2_{id}$, $\sigma^2_{se}$, and $\sigma^2_{ie}$. In the numerical SRT evaluation, the fading amplitudes are first generated by using the exponential distribution having different means for different wireless channels, which are then substituted into the specific definition of an outage (or intercept) event for determining the OP (or IP). In our computer simulations, the means of the squared fading magnitudes are set to $\sigma^2_{sd}=1$, $\sigma^2_{si}=\sigma^2_{id}=2$, and $\sigma^2_{se}=\sigma^2_{ie}=0.2$. {{It needs to be pointed out that although only the Rayleigh fading is considered in this article for the numerical SRT evaluation, similar SRT results can be obtained for other fading models e.g. Nakagami and Rice fading.}} Additionally, {{an SNR of $15{\textrm{ dB}}$}} is used in the numerical SRT evaluations.

\begin{figure}
  \centering
  {\includegraphics[scale=0.8]{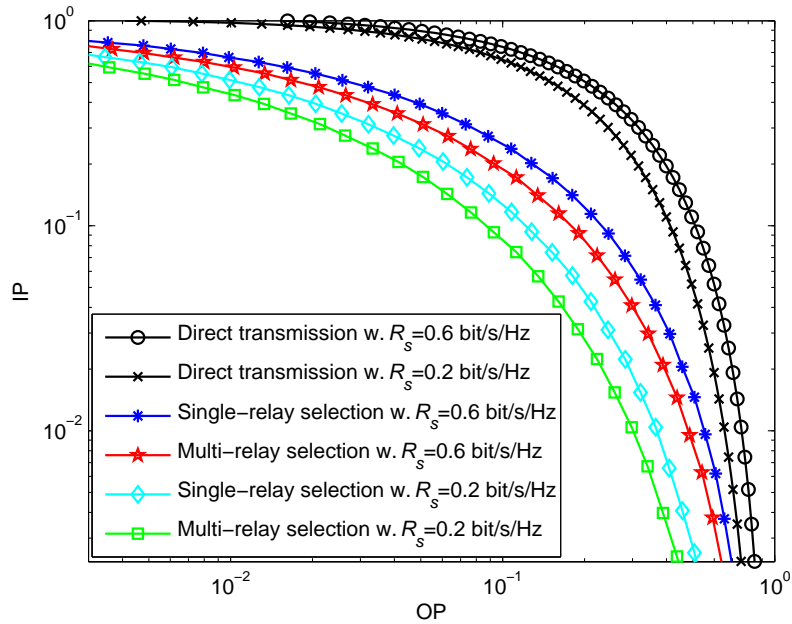}\\
  \caption{IP versus OP of the direct transmission as well as of the SRS and MRS schemes for different secrecy rates in conjunction with {{an SNR of $15{\textrm{dB}}$}}, $N=4$, $\sigma^2_{sd}=1$, $\sigma^2_{si}=\sigma^2_{id}=2$, and $\sigma^2_{se}=\sigma^2_{ie}=0.2$.}\label{Fig5}}
\end{figure}
Fig. 5 shows the IP versus OP of the direct transmission (DT) as well as the SRS and MRS schemes for different secrecy rates associated with $N=4$, where $N$ is the number of RNs. It can be observed from Fig. 5 that for the cases of $R_s=0.2{\textrm{ bit/s/Hz}}$ and $0.6{\textrm{ bit/s/Hz}}$, both the SRS and MRS schemes outperform the conventional DT in terms of their SRTs. Moreover, the SRT of MRS is better than that of SRS, explicitly showing the advantages of multi-relay selection over single-relay selection. It is worth mentioning that the security benefit of the MRS over SRS is achieved at the expense of a higher implementation complexity, since the MRS requires the complex symbol-level synchronization among multiple spatially-distributed RNs, compared to the SRS. Fig. 5 also illustrates that given a specific target OP, the IPs of the DT, SRS and MRS schemes all increase, as the secrecy rate increases from $R_s=0.2{\textrm{ bit/s/Hz}}$ to $0.6{\textrm{ bit/s/Hz}}$.

\begin{figure}
  \centering
  {\includegraphics[scale=0.8]{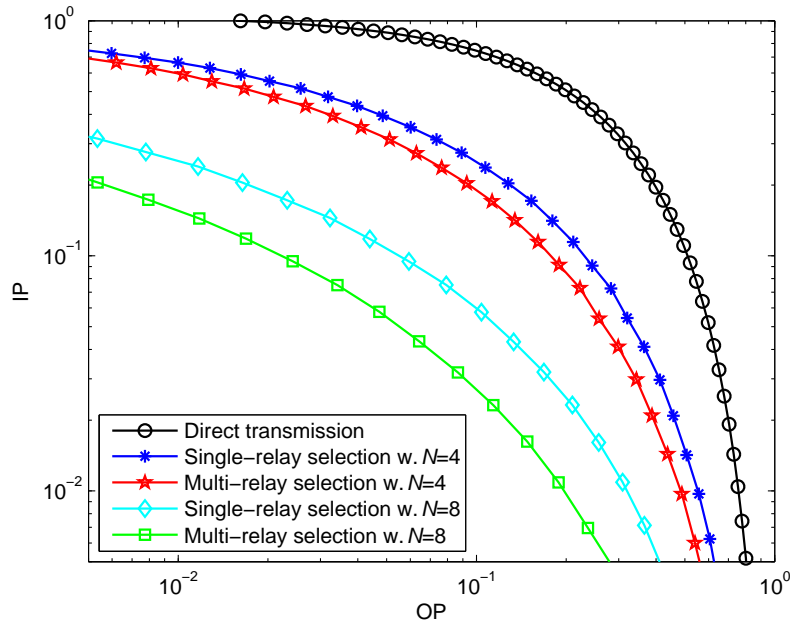}\\
  \caption{IP versus OP of the direct transmission as well as of the SRS and MRS schemes for different number of relays $N$ in conjunction with {{an SNR of $15{\textrm{ dB}}$}}, $R_s=0.6{\textrm{ bit/s/Hz}}$, $\sigma^2_{sd}=1$, $\sigma^2_{si}=\sigma^2_{id}=2$, and $\sigma^2_{se}=\sigma^2_{ie}=0.2$.}\label{Fig6}}
\end{figure}
In Fig. 6, we characterize the IP versus OP of the DT as well as the SRS and MRS schemes for different number of relays $N$. As shown in Fig. 6, the MRS scheme is the best and the conventional DT performs the worst in terms of their SRTs, which further confirms the security advantage of exploiting the multi-relay selection for guarding against eavesdropping. It is also observed from Fig. 6 that as the number of relays $N$ increases from $N=4$ to $8$, the SRTs of both the SRS and MRS schemes improve significantly. This implies that increasing the number of relays is an effective means of enhancing the level of wireless physical-layer security against eavesdropping attack.

\section{Open Challenges and Future Work}
This section is mainly focused on the discussion of open issues in the PLS of wireless relay networks. Although recently extensive efforts have been devoted to this research subject, many challenging issues still remain open at the time of writing.

\subsection{Wireless Security with Untrusted Relays}
As discussed above, the IP of wireless communications relying on relay selection can be significantly reduced upon increasing the number of relays. It has to be pointed out that in both the SRS and MRS schemes, the relays are assumed to be trusted without any intention of tapping the legitimate transmissions. However, this assumption may be invalid in some cases, where the relays are compromised and become untrusted. Hence, it is of importance to explore whether an untrusted relay is still beneficial or not in terms of enhancing the PLS. At the time of writing, physical-layer network coding is considered to be an attractive design alternative for guarding against an untrusted relay, where a pair of transceivers are allowed to transmit simultaneously so that the relay only receives a composite superimposed or mixed signal. Although physical-layer network coding improves the security level, an untrusted relay can still decode its composite signal by using successive interference cancellation techniques. It is challenging, but interesting to examine the benefits of relay selection in terms of improving the wireless PLS, while keeping the legitimate transmission completely confidential to the relays.

\subsection{Joint Multi-Relay and Multi-Jammer Selection}
In wireless relay networks, a relay node can either be used to assist the source transmission for improving the quality of the legitimate channel, or to act as a jammer for imposing artificial interference on an eavesdropper for deliberately contaminating the wiretap channel. When multiple RNs are available, some relays can be carefully selected for enhancing the throughput of the main channel, while others may be used as jamming nodes for interfering with the eavesdropper. This scenario is referred to as joint relay and jammer selection, which may be invoked for improving the wireless security in the face of eavesdropping attacks. Existing research efforts are limited to the single-relay and single-jammer selection, which could be further extended to a more general framework of multi-relay and multi-jammer selection. Additionally, typically perfect CSI knowledge of the main and/or wiretap links is assumed in literature, which is not practical due to the presence of CSI estimation errors. It will be of interest to investigate joint multi-relay and multi-jammer selection in the face of imperfect CSI knowledge of the main and/or wiretap links.

\subsection{Location-Aware Relaying against Eavesdroppers}
Having location information is crucial for determining whether or not a RN is helpful for assisting the legitimate S-D transmission against an eavesdropper. For example, if the RN is much closer to both S and D than to E, it would achieve significant security benefits when employing the RN for forwarding the source signal to D. On the other hand, if the RN happens to be located closely to E, exploiting relay aided transmission may not achieve beneficial security improvements. {{This implies that the deployed network environment (e.g. the positions of RNs) would have an impact on the SRT performance of wireless relay networks, which is to be addressed in the future. Moreover, only the single eavesdropper is considered in this article for performing the relay selection, but a more general wireless network may consist of multiple eavesdroppers. Therefore, it is important to determine where the RNs should be deployed in a certain geographic area for the sake of maximizing the wireless PLS in the presence of multiple eavesdroppers}}, which is an interesting open challenge for the future.

\section{Concluding Remarks}
In this article, we studied the benefits of relay selection from a SRT perspective in wireless networks in the presence of an eavesdropper, where multiple RNs are available for protecting the S-D transmission against eavesdropping. We presented a pair of relay selection schemes, namely the SRS and MRS, where the SRS only selects a single ``best" RN to assist the legitimate transmission from S to D, whereas in the MRS scheme, multiple relays are allowed to simultaneously forward the source transmission. Numerical SRT results were provided for characterizing the performance comparison among the DT, SRS and MRS schemes in terms of their IP and OP. It was shown that the MRS scheme achieves the best SRT and the DT method performs the worst. As the number of RNs increases, the SRTs of both the SRS and MRS schemes improve accordingly, explicitly showing the security advantage of exploiting relay selection. Finally, we pointed out some open challenges in the field of PLS of wireless relay networks, including the untrusted relay issues, joint multi-relay and multi-jammer selection as well as the location-aware relaying against eavesdropping.

\section{Acknowledgement}
This work was supported by the National Natural Science Foundation of China (Grant Nos. 61401223 and 61522109), the Natural Science Foundation of Jiangsu Province (Grant Nos. BK20140887 and BK20150040), and the Key Project of Natural Science Research of Higher Education Institutions of Jiangsu Province (No. 15KJA510003).

\clearpage

\section{Biographies}

\textbf{Yulong Zou} (SM'13) is a Professor at the Nanjing University of Posts and Telecommunications (NUPT), Nanjing, China. He received the B.Eng. degree in Information Engineering from NUPT, Nanjing, China, in July 2006, the first Ph.D. degree in Electrical Engineering from the Stevens Institute of Technology, New Jersey, USA, in May 2012, and the second Ph.D. degree in Signal and Information Processing from NUPT, Nanjing, China, in July 2012. (Email: yulong.zou@njupt.edu.cn)\\

\textbf{Jia Zhu} is an Associate Professor at the Nanjing University of Posts and Telecommunications (NUPT), Nanjing, China. She received the B.Eng. degree in Computer Science and Technology from the Hohai University, Nanjing, China, in July 2005, and the Ph.D. degree in Signal and Information Processing from NUPT, Nanjing, China, in April 2010. (Email: jiazhu@njupt.edu.cn)\\

\textbf{Xuelong Li} (M'02-SM'07-F'12) is a Professor with the Center for OPTical IMagery Analysis and Learning (OPTIMAL), State Key Laboratory of Transient Optics and Photonics, Xi'an Institute of Optics and Precision Mechanics, Chinese Academy of Sciences, Xi'an, Shaanxi, China. (Email: xuelong\_li@opt.ac.cn)\\

\textbf{Lajos Hanzo} is a Chair Professor with the School of Electronics and Computer Science, University of Southampton, Southampton, United Kingdom. He received his degree in electronics in 1976 and his doctorate in 1983. In 2009 he was awarded ``Doctor Honoris Causa" by the Technical University of Budapest. (Email: lh@ecs.soton.ac.uk)\\


\begin{thebibliography}{11}

\bibitem{IEEEhowto:1}
R. Bassily, E. Ekrem, X. He, \emph{et al.}, ``Cooperative security at the physical layer: A summary of recent advances," \emph{IEEE Sig. Process. Mag.}, vol. 30, no. 5, May 2013, pp. 16-28.

\bibitem{IEEEhowto:2}
S. Wang, X. Xu, and W. Yang, ``Physical layer security in underlay CCRNs with fixed transmit power," \emph{KSII Trans. Internet and Inform. Syst.}, vol. 9, no. 1, Jan. 2015, pp. 260-279.

\bibitem{IEEEhowto:3}
H. Chen, Y. Cai, and D. Wu, ``Joint spectrum and power allocation for green D2D communication with physical layer security consideration," \emph{KSII Trans. Internet and Inform. Syst.}, vol. 9, no. 3, Mar. 2015, pp. 1057-1073.

\bibitem{IEEEhowto:4}
A. D. Wyner, ``The wire-tap channel," \emph{Bell Syst. Tech. Journal}, vol. 54, no. 8, Aug. 1975, pp. 1355-1387.

\bibitem{IEEEhowto:5}
S. K. Leung-Yan-Cheong and M. E. Hellman, ``The Gaussian wiretap channel," \emph{IEEE Trans. Inf. Theory}, vol. 24, no. 7, Jul. 1978, pp. 451-456.

\bibitem{IEEEhowto:6}
Y. Zou, J. Zhu, X. Wang, and V. Leung, ``Improving physical-layer security in wireless communications through diversity techniques," \emph{IEEE Net. Mag.}, vol. 29, no. 1, Jan. 2015, pp. 42-48.

\bibitem{IEEEhowto:7}
S. Goel and R. Negi, ``Guaranteeing secrecy using artificial noise," \emph{IEEE Trans. Wireless Commun.}, vol. 7, no. 6, Jul. 2008, pp. 2180-2189.

\bibitem{IEEEhowto:8}
A. Mukherjee amd A. Swindlehurst, ``Robust beamforming for security in MIMO wiretap channels with imperfect CSI," \emph{IEEE
Trans. Signal Process.}, vol. 59, no. 1, Jan. 2011, pp. 351-361.

\bibitem{IEEEhowto:9}
F. Zhu, F. Gao, M. Yao, and H. Zou, ``Joint information- and jamming-beamforming for physical layer security with full duplex base station," \emph{IEEE Trans. Signal Process.}, vol. 62, no. 24, Dec. 2014, pp. 6391-6401.

\bibitem{IEEEhowto:10}
Y. Zou, X. Wang, and W. Shen, ``Optimal relay selection for physical-layer security in cooperative wireless networks," \emph{IEEE J. Select. Areas Commun.}, vol. 31, no. 10, Oct. 2013, pp. 2099-2111.

\bibitem{IEEEhowto:11}
Y. Zou, J. Zhu, L. Yang, Y.-.C. Liang, and Y.-D. Yao, ``Securing physical-layer communications for cognitive radio networks," \emph{IEEE Commun. Mag.}, vol. 53, no. 9, Sept. 2015.

\bibitem{IEEEhowto:12}
Y. Zou, X. Wang, W. Shen, and L. Hanzo, ``Security versus reliability analysis of opportunistic relaying," \emph{IEEE Trans. Veh. Tech.}, vol. 63, no. 6, Jul. 2014, pp. 2653-2661.

\bibitem{IEEEhowto:13}
O. Koyluoglu and H. Gamal, ``Polar coding for secure transmission and key agreement," \emph{IEEE Trans. Inf. Foren. Sec.}, vol. 7, no. 5, Oct. 2012, pp. 1472-1483.

\bibitem{IEEEhowto:14}
Y. Xiao, H.-H. Chen, B. Sun, R. Wang, and S. Sethi, ``MAC security and security overhead analysis in the IEEE 802.15.4 wireless sensor networks," \emph{EURASIP J. Wirel. Commun. and Net.}, DOI:10.1155/WCN/2006/93830, 2006.

\bibitem{IEEEhowto:15}
X. Tang, R. Liu, P. Spasojevic, and H. V. Poor, ``On the throughput of secure hybrid-ARQ protocols for Gaussian block fading channels," \emph{IEEE Trans. Inf. Theory}, vol. 55, no. 4, Apr. 2009, pp. 1575-1591.

\end{thebibliography}
\end{document}